# The $k$-space Origins of Scattering in $Bi_2Sr_2CaCu_2O_{8+x}$

Jacob W. Alldredge[1], Eduardo M. Calleja[1], Jixia Dai[1], H. Eisaki[2], S. Uchida[3], Kyle McElroy[1]

1 University of Colorado Boulder, Boulder CO

2 Institute of Advanced Industrial Science and Technology, Tsukuba, Ibaraki 305-8568, Japan

3 Department of Physics, University of Tokyo, Bunkyo-ku, Tokyo 113-0033, Japan

## Abstract:

We demonstrate a general, computer automated procedure that inverts the $q$-space scattering data measured by spectroscopic imaging scanning tunneling microscopy (**SI-STM**) to determine the $k$-space scattering structure. This allows a detailed examination of the $k$-space origins of the quasiparticle interference (**QPI**) pattern in $Bi_2Sr_2CaCu_2O_{8+x}$. This new method allows the measurements of the differences between the positive and negative energy dispersions, the gap structure and it also measures energy dependent scattering length scale. Furthermore, the transitions between the dispersive QPI, the checkerboard and the pseudogap are mapped in detail allowing the exact nature of these transitions to be determined for both positive and negative energies. We are also able to measure the $k$-space scattering structure over a wide range of doping ($p \sim 0.22$ to 0.08), including regions where the octet model is not applicable. Our technique allows a complete picture of the $k$-space origins of the spatial excitations in $Bi_2Sr_2CaCu_2O_{8+x}$ to be mapped out, providing for better comparisons between SI-STM and other experimental probes of the band structure and validating our new general approach for determining the $k$-space scattering origins from SI-STM data.

## Introduction:

One advantage of SI-STM is its ability to resolve the interference pattern in the density of states caused by the scattering of quasiparticles in a crystal[1-12]. The scattered quasiparticles wave vectors contain information not only about the scattering process, but also about the underlying band structure of the crystal. However, so far the analysis of such scattering patterns has been limited due to the difficulty in inverting the SI-STM data in order to determine the $k$-space scattering structure[11-15]. In order to solve this problem, we have developed a constrained Monte Carlo reconstruction (**CMCR**) method of determining the $k$-space sources of the scattering. Our method is based on the joint density of states (**JDOS**)[15] model and provides a basic method that can be easily modified to take into account additional information concerning the scattering, such as scattering type, number of bands and selection rules. We have aimed to make this technique as general as possible in order to allow its application to any QPI data sets collected by SI-STM. In this paper we apply the CMCR method to the widely studied $Bi_2Sr_2CaCu_2O_{8+x}$ (**Bi-2212**) data sets in order to validate it and make a more direct comparison between SI-STM and angle-resolved photo emission spectroscopy (**ARPES**). This allows us to resolve ongoing questions about gap structure, the influence of the parent compound anti-ferromagnetic zone boundary (**PAF-zone**) and doping evolution of the $k$-space scattering structure.

By applying the CMCR method to the QPI in in $Bi_2Sr_2CaCu_2O_{8+x}$, we are able to reconstruct the $k$-space scattering structure for positive and negative energies separately and measure an energy dependent effective scattering length scale. This scattering length scale shows the transition between the low energy dispersive QPI, the $q_1^*$ excitation (previously associated with the checkerboard) and the pseudogap state[5,6,16]. We are also able to measure the $k$-space structure over a large doping range and in a range where the QPI no longer consists of 7 independent $q$-vectors. The completeness of the reconstruction by the CMCR method allows the fitting of a tight binding band structure, and the reconnection can be matched to ARPES data. The reconstruction allows for a more complete measurement of the gap structure, which allows the SI-STM measurement of a higher harmonic component[1,4,16] to be reconciled with ARPES lowest order d-wave gap measurements[17]. The PAF-zone boundary's influence on the scattering is resolved across all dopings and its energy dependence is shown. This energy dependence reveals that the PAF-zone boundary has an influence on the QPI at energies much smaller than the its termination energy[18]. Thus with the application of the CMCR method we are able to bet-ter understand the scattering phenomena and its signature in the SI-STM data. This is important for determining the interaction of the scattering with the phenomena present in cuprates and how in determining how these phenomena relate to the pairing mechanism.

## $Bi_2Sr_2CaCu_2O_{8+x}$ QPI Background:

The QPI patterns measured by SI-STM are due to large scale scattering of quasiparticles by the many weak scatters present in the material[19–25]. This is the same mechanism as Friedel oscillations; however in Bi-2212 and other materials, the scatterer itself is not imaged due to the lack of a clearer scatterer resonance. The resulting QPI pattern, while displaying large scale intensity variations, does not appear to show the traditional exponential decay of Friedel oscillations, due to the large number of scatters[23–28], the coherent nature of the quasiparticles and/or the extend states of the scatters[19,29–31]. By taking the Fourier transform (**FT**) of the QPI, the interference of the quasiparticle waves can be resolved in *q*-space. Theoretically this pattern can be understood by using the spectral density of states, A(*k*,ω), which represents the *k*-space density of states[32,33]. When it is combined with Fermi's Golden Rule, it determines the *q*-vectors of the quasiparticles whose intensities are proportional to the number of states at the origin and end of the q-vector. That is the quasiparticles predominantly scatter between the sections of the *k*-space density of states that have the highest number of states. A simple method of determining these *q*-vectors is accomplished by calculating the JDOS for a given A(*k*,ω). A sample JDOS and *k*-space density of states is shown in figure 1. The JDOS is a measure of all the possible scattering vectors and places with a larger number of states in *k*-space will act as a large sources or sinks for the scattering vectors[13] giving rise to intense *q*-vectors connecting these points in the JDOS. Practically, the JDOS can be calculated by taking the autocorrelation[13,15] of the A(*k*,ω) over an appropriate *k*-space range.

The JDOS analysis technique has been used with ARPES measurements of the spectral density of states to reproduce the SI-STM observed QPI patterns with some degree of accuracy[15,34–36]. However, the ARPES based JDOS calculations tend to produce spatially large, connected QPI patterns and not the coalesced peaks seen by SI-STM in Bi-2212. This is thought to be caused by either a matrix element or an effect due to the nature of the scattering that causes the QPI[15]. The differences between types of scattering can be theoretically modeled by using the τ matrices for different scatters combined with a superconducting Green's function or A(*k*,ω)[19]. This approach has been validated in samples with vortex cores where the *q*-vectors intensities change in a manner that is consistent with an increase in the magnetic scattering[37] or τ 2 scattering.

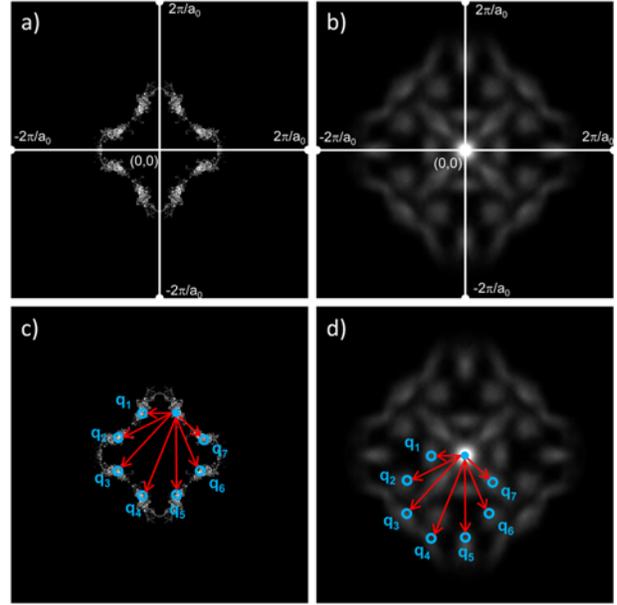

**Figure 1:** Using the -8 meV layer of our reconstruction for UD74 K we demonstrate how the JDOS produces the QPI pattern and its relation to the octet model. a) The -8 meV reconstructed *k*-space pattern. This represents the *k*-space states that contribute to the scattering observed in the QPI pattern. By taking the autocorrelation of the reconstruction we can generate a JDOS, which is then compared to the data. b) The JDOS derived from the autocorrelation of a). c) The eight points that are used in the 'octet model'. By taking these points that are represented by the centers of the *k*-space structure shown in c) we can show that these same vectors when translated over to *q*-space in d) represent the QPI vectors recorded by the SI-STM.

In Bi-2212 the QPI pattern is made up of 7 *q*-vectors with 4-fold symmetry[1] (see figure 1c,d). This distinctive pattern's *k*-space origins have been explained by the octet model[13] which uses the superconducting band structure and Fermi's Golden Rule to determine that the scattering will occur between the eight ends of the superconducting bands. This is the same result as the JDOS method, if one assumes that the spectral density of states consists of only the points at the ends of the superconducting band structure[4]. These 7 *q*-vectors disperse and by mapping them out as a function of energy and using the octet model, the underlying band structure and superconducting gap can be determined[5,6,13,38]. This measurement, carried out as a function of doping, shows the band structures shifting outwards with doping in accordance with the Luttinger theory for Mott insulators[5,6,38,39] and that the superconducting gap is not a pure d-wave gap, but has an additional higher harmonic Cos[6*θ] contribution that increases with decreasing doping[5,6,18,38]. This gap structure is in contrast with any of the different gap structures measured by ARPES[17,40–42].

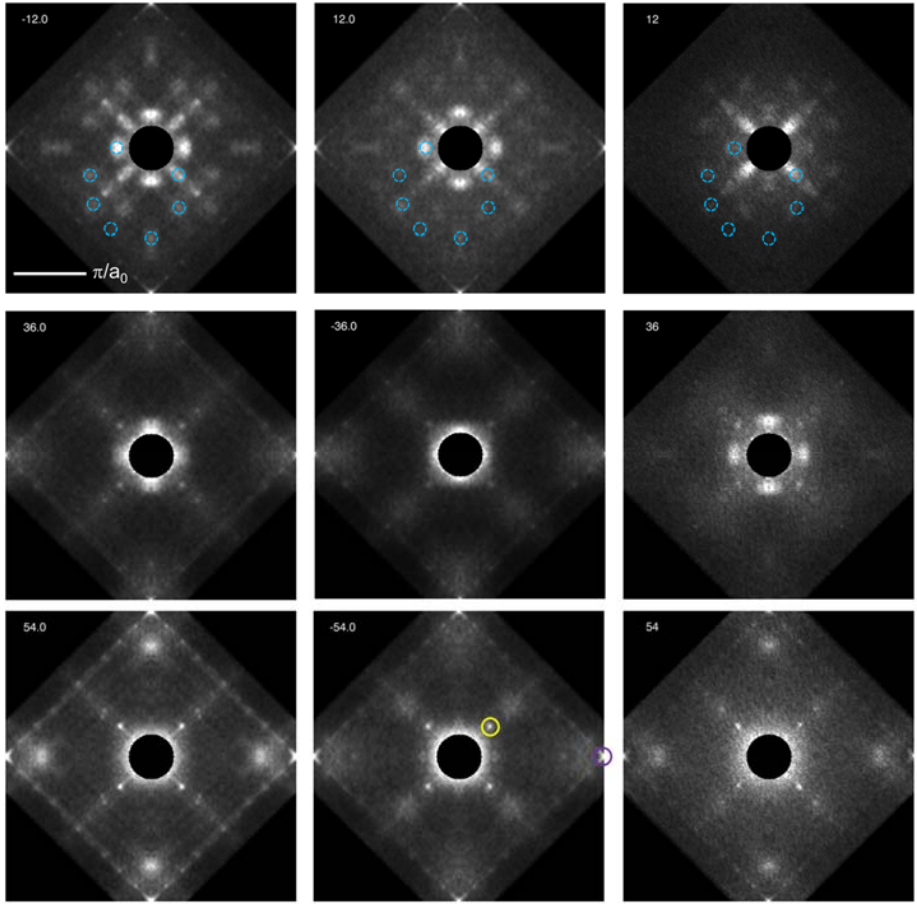

**Figure 2:** The three different energy regimes spatial excitations, UD74 K data used as an example. Left column is negative energies, middle column positive energies and right column is the ratio map. In the top row, at low energies, represented here by 12 meV, we have the dispersive QPI pattern of 7 4-fold symmetric peaks (circled with the blue dotted lines). At higher energies, middle row, the dispersive QPI disappears and is replaced on the positive energy side with the $q_1^*$ excitation. We show 36 meV energy layer here for this example. The $q_1^*$ excitation is absent in the negative column and strong in the positive. At energies around the gap value the pseudogap spatial excitation is observed, shown here in the bottom row using the 54 meV energy layer. This $S_{x,y}$ peak at ~ 2 (3/4 $\pi/a_0$) is seen strongly in the ratio map and at positive energies. The purple circle marks the atomic peak and the yellow circle marks the supermodulation peak.

The dispersive QPI pattern has also been shown to 'terminate' when the octet models determined $k$-space points reach the PAF-zone boundary[18], or the line in $k$-space from ($\pi/a_0$,0) to (0,$\pi/a_0$). When the dispersing QPI pattern reaches this line, the signal to noise ratio becomes too low to determine the $k$-space points using the octet model[38]. The QPI is also replaced with two non-dispersing peaks, $q_1^*$ and $q_5^*$ whose wave vectors are determined by the intersection of the band structure and the PAF-zone boundary[5,6]. $q_5^*$ has been shown to be smectic and can be divided into two intensities depending on the direction resulting in two different vectors $S_x$ and $S_y$. It also has a peak in intensity at the LDOS(E) measured gap value and this gap, and the $q$-space peak in intensity, scale with the pseudogap energy[6,43]. $q_1^*$ exists in a region between the low energy dispersive QPI states and the pseudogap states[16,44]. It coincides in energy with the kink in the LDOS(E) that exists between the low energy homogeneous V shaped states and the high energy peak[16,44,45].

Previous studies of the QPI have focused on analyzing the ratio maps QPI pattern[5,8,9]. The ratio map is the FT of the ratio of the positive energy states to the negative energy states and is thought to eliminate some of the setup effects present in SI-STM[46]. These setup effects cause modulations that are present on one side of zero energy to be mirrored and spread across energy on the opposite side[16]. However, the ratio map QPI also eliminates the differences in the negative and positive dispersions and looks instead at the overlap between the wave vectors. Figure 2 shows example Bi-2212 QPI patterns for the three energy scales, the dispersive QPI at low energies, the $q_1^*$ excitation at intermediate energies and the pseudogap at higher energies. This data is taken from the UD74 K data and the energy span of the checkerboard at this doping is very small and overlaps with the QPI and pseudogap in energy[16,44]. In figure 2 top row, it is clear that the positive and negative $q$-vectors at the same energy have different wavelengths. The resulting ratio map at that energy also has the effect of removing $q_1/q_1^*$ and $q_5$. The removal of the strong $q_1/q_1^*$ vector is thought to be largely because of its increased presence due to the setup effect[16,18], which is removed by the ratio map. The disappearance of $q_5$ is thought to have to do with it having a difference in phase between the positive and negative energies[47]. In figure 2 middle row, the $q_1^*$ modulation is highlighted. At this high doping it is mainly seen in the ratio map in the right hand column. The bottom row shows the $S_{x,y}$ modulation which seems to appear only at positive energies[16].

Previous studies of the QPI patterns in Bi-2212 have drawn comparisons between theoretical JDOS or ARPES measurements and the FT SI-STM data[15,48] or have used the octet model to extract the $\mathbf{k}$-space origins of the QPI data[1,4–6,8,18,38]. Below we present a new method that uses the JDOS framework to determine the $\mathbf{k}$-space scattering directly from the data, with no other theoretical constraints or models. This provides a more direct method of measuring the $\mathbf{k}$-space scattering structure and one that can be applied to materials with unknown band structures and scattering in the future. It also provides a framework that allows the measure of the $\mathbf{k}$-space dispersion as it transitions to the $q_1^*$ excitation at the PAF-zone boundary.

## Method:

Our CMCR based inversion method draws inspiration from several different inverse methods that are aimed at inverting incomplete Fourier space data in order to recover the real space image[14,49–52]. The method aimed at inverting QPI specifically[14], uses the JDOS to go from $\mathbf{k}$-space to $\mathbf{q}$-space and while we have found that this method works when used to invert a test JDOS based on a theoretical $A(\mathbf{k},\omega)$ generated with a superconducting gap, it fails when applied to the SI-STM QPI data or binary masks of the data designed to negate the QPI peaks intensity fluctuations. Not only does it fail, but the other general methods all fail, similarly, when suing the JDOS model. The exact reason why these methods fail is unknown, it could be due to the noise levels of the actual data, the addition of non-QPI related $\mathbf{q}$-space peaks (atomic lattice, supermodulation) or matrix element/scattering selection effects.

CMCR uses the JDOS to generate a QPI pattern from a $\mathbf{k}$-space test pattern. The JDOS method has been used in the development of the octet model[13], QPI studies using $\mathbf{k}$-space ARPES data[15,48] and in attempts to explain the QPI pattern in $Sr_3Ru_2O_7$. This last study[12] of $Sr_3Ru_2O_7$ involved making educated guesses about the band structure involved in the scattering and using the JDOS to generate test QPI patterns. Then the band structure guess was refined by hand until the test QPI pattern was matched to the QPI data. This process served as the inspiration for this work and here we have replaced this educated guess work in determining the $\mathbf{k}$-space scattering points with an automated procedure which allows for a computer determined band structure with minimal human intervention. This removes the man hours needed to back out a model $\mathbf{k}$-space scattering and allows for a more accurate determination of the $\mathbf{k}$-space structure to be made. The CMCR method is also highly expandable. For Bi-2212, we have a compound that can be modeled with a simple symmetric one-band model; however the CMCR method can easily be expanded to include multiple bands, matrix elements and scattering selection rules, allowing for a thorough investigation of the scattering processes.

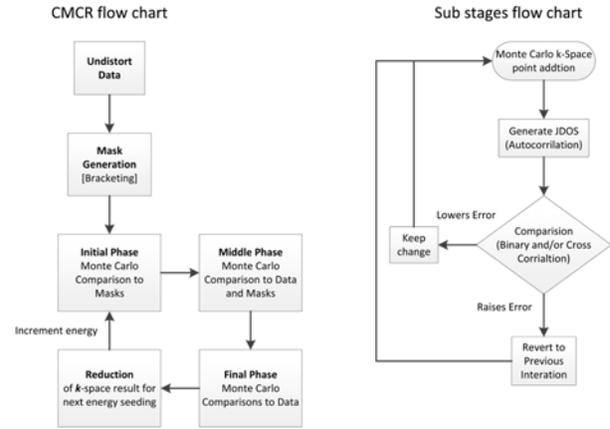

**Figure 3:** Algorithm flowchart. The left-hand side outlines the 6 major sections of the $\mathbf{k}$-space reconstruction process. On the right is the flowchart for each of the Monte Carlo stages. The key difference between the 4 phases is if they make the comparison to the data, the mask or both and how they determine the error.

## Procedure:

Figure 3 shows a flowchart of the CMCR process. It begins by properly setting up the data, removing skewing distortions by setting the atomic lattice peaks to be at $2\pi/a_0$. The data is then symmetrized across the 4 quadrants since the electronic structure is 4-fold symmetric. A set of binary masks are then generated from the data. These masks are used to highlight the QPI structures and provide a binary selection of the areas in $\mathbf{q}$-space which are important and also to provide a $\mathbf{q}$-space image without the intensity variations of the different $\mathbf{q}$-vectors. The masks also help to constrain the CMCR algorithm and three different ones are generated for each of the energies in the QPI data. The amount of area in each mask is varied in order to bracket the data, high and low.

The initial phase of the CMCR algorithm generates a binary $\mathbf{k}$-space structure for each of the masks. This is done by randomly selecting points in $\mathbf{k}$-space, generating a test QPI structure and comparing the mask to the test QPI structure. If the addition of the points reduces the error, then the points are kept. If the addition of the points does not reduce the error, then another iteration is attempted. The error is computed by determining the points there test QPI and test QPI pattern overlap and this is weighted by a term that accounts for the overall size of the $k$-space structure. The size term is included in order to ensure the smallest possible $k$-space structure for a given QPI pattern. The mask also acts as a noise level cutoff, by restricting the total expanse of the $k$-space structure the CMCR algorithm is prohibited from filling every point in $k$-space to mimic the background noise. The test $k$-space structure has enforced 8-fold symmetry in this phase since the data is close to 8 fold symmetric and this speeds up the process.

After the binary masks are used to generate a binary, restricted size, $k$-space structure, the middle phase is begun. The middle phase uses the $k$-space test structure of the initial phase, but adds direct comparison with the data to the error calculation along with the mask error and the $k$-space size weight. The data comparison error is measured by calculating the normalized cross correlation between the test QPI pattern and the QPI data. This phase allows the reconstruction of the intensity of the QPI, although it is slower due to the cross correlation calculation.

The final phase uses only the normalized cross correlation between the data and the test QPI as a measure of the error in the reconstruction. It is further constrained by only allowing the addition and subtraction of points that already have weight in $k$-space. This prevents the unconstrained growth of the solution and further refines the reconstruction.

When the CMCR process is completed for a given energy layers solution is compared to the next energies binary masks. The solution is then trimmed down in area to provide a seed for the next energies reconstruction. This is accomplished by taking the current energies solution and removing points from it that produce features that are not present in the next energies mask. This phase assumes that the $k$-space dispersion is continuous in energy and it ensures a continuous dispersion in the solution, while speeding up the overall process. When the process is complete, all three $k$-space reconstructions (one for each of the three masks) are added together to form the final reconstruction.

**Length scale masks:**

The JDOS method of generating a QPI pattern cannot account for the variation in the intensities of the QPI vectors that is seen in the data. This is because the JDOS method gives a weight to each $q$-vector only through the number of states at its two corresponding $k$-space points and it has no method of decreasing the intensity of the peaks based off their wavelength. The data has a pronounced intensity variation where the long wavelength $q$-vectors are always more intense than the shorter wavelength ones. In order to account for this variation and to track it as a function of energy, we introduce a simple Gaussian mask that decreases the intensity of the shorter wavelength vectors. This allows the accurate reproduction of the QPI pattern and the determination of an effective length scale as a function of energy. This is the simplest model that can reproduce the data and it can easily be replaced in the future with more complex scattering selection rules that, given enough processing power, can be added into the CMCR algorithm.

**Results:**

Due to the extended 3D nature of the data it is rather hard to present it all within the confines of a 2D journal article. Hence we show an example of the reconstructions individual energy layers for the UD74 K data and how it compares with the data. Then the detail of one of the 4-fold symmetric quadrants is shown, allowing a new perspective on the higher harmonic gap structure. By integrating the $k$-space reconstruction over energy, we can make comparisons to ARPES, examine the positive and negative dispersions and present the doping dependence of the $k$-space reconstruction in a compact form.

**Reproduction of the $q$-space pattern:**

Figures 4 and 5 provide examples of the $k$-space reconstruction and the corresponding $q$-space reconstructions for the UD74 K data's negative energies. In the leftmost column is the $k$-space reconstruction and the JDOS reconstruction generated from this pattern is in the next column over. In order to reproduce the observed data, shown in the right most column, a Gaussian mask is fit to each energy with only its full width half maximum (**FWHM**) being varied. The $q$-space reconstruction multiplied by the fit Gaussian mask is found in the $3^{rd}$ column from the left. These figures show that the CMCR method has the ability to reproduce the data over a wide energy range. It also shows the dispersion of the $k$-space structure outward from the node to the antinode with increasing energy. The curving inward of the $k$-space reconstruction when it reaches the PAF-zone boundary evident at the antinode in each reconstruction. Even at energies lower than the termination energy (~ 34 meV for this data set[18,38]) the $k$-space reconstruction captures the effect of the PAF-zone boundary and shows its influence at energies much lower than the termination energy. At zero energy we can see a departure in the data from the octet QPI pattern caused by the zinc impurities

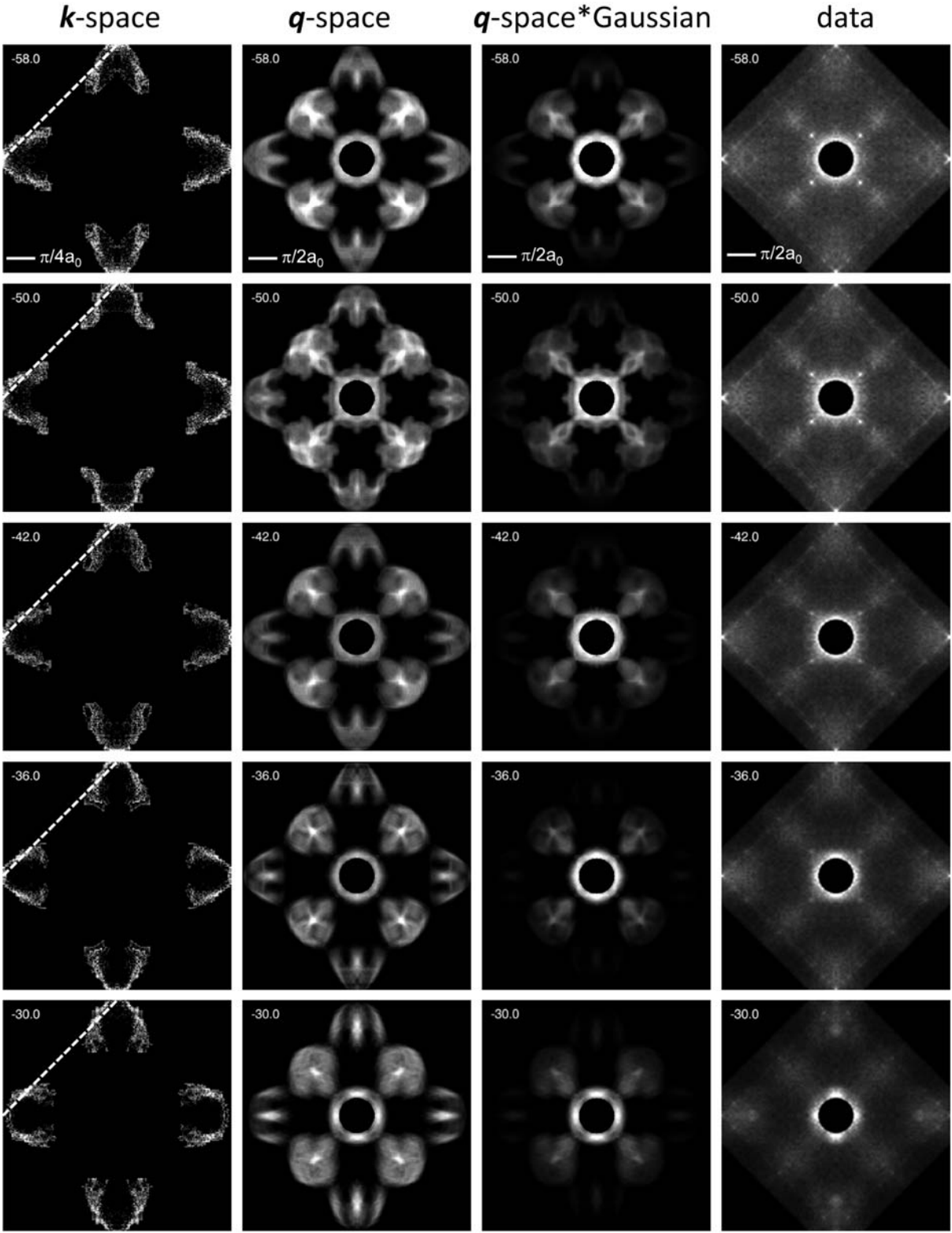

**Figure 4:** The *k*-space reconstruction, the QPI autocorrelation reconstruction, the QPI autocorrelation reconstruction Gaussian masked and the symmetrized data. The white dotted line represents the PAF-zone boundary in one quadrant.

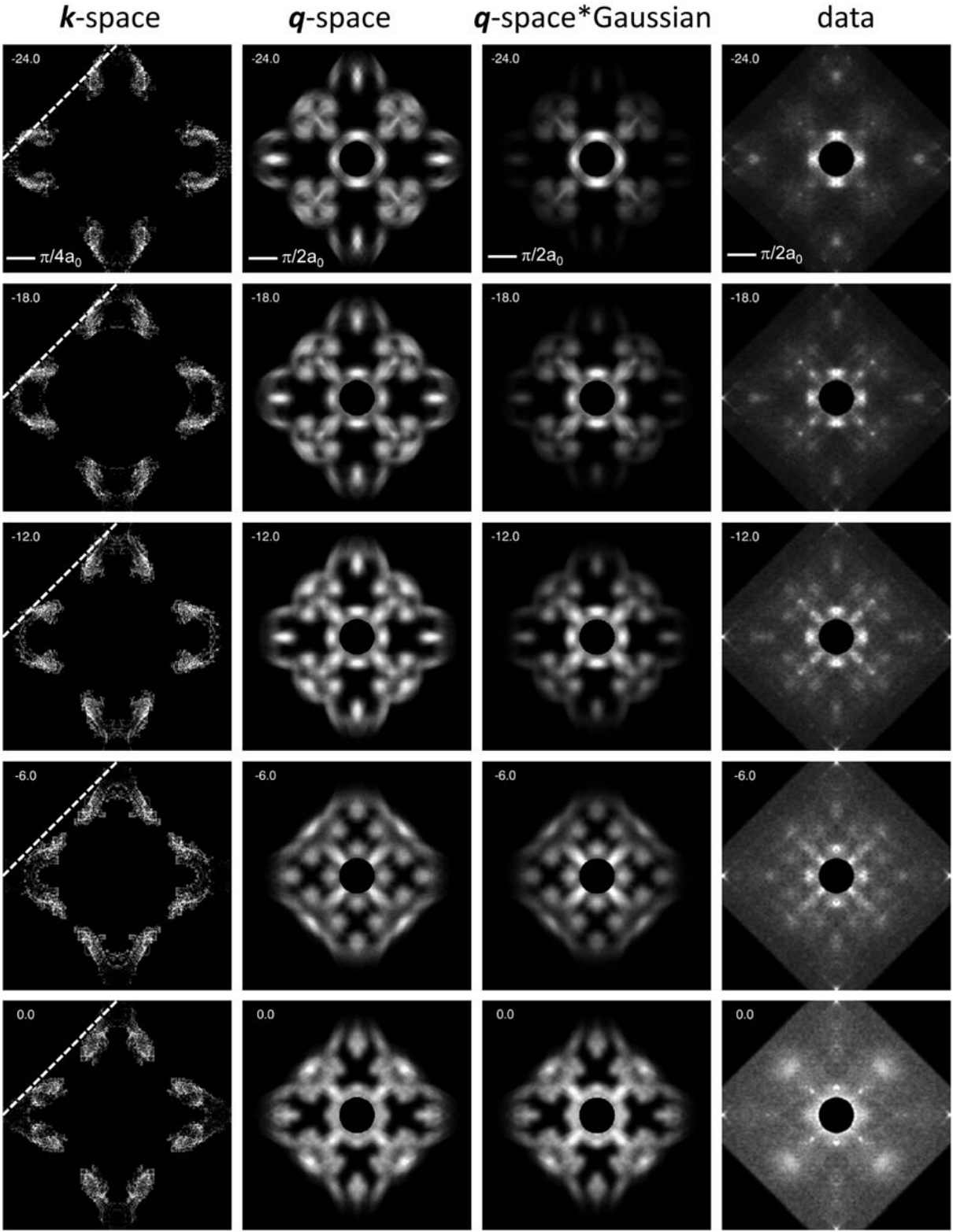

**Figure 5:** Continuation of the **k**-space reconstruction, the QPI autocorrelation reconstruction, the QPI autocorrelation reconstruction Gaussian masked and the symmetrized data. The white dotted line represents the PAF-zone boundary in one quadrant.

and copper site vacancies or scattering symmetry considerations[7,53].

Figure 6a is the normalized cross correlation between the $q$-space reconstruction and the data for the UD74 K data set shown in figures 4 and 5. This demonstrates that the reconstruction has a high degree of correlation with the data over the whole energy range. At low energies we see a decreasing normalized cross correlation as the strength of the QPI pattern decreases. The lower normalized cross correlation at low energies is likely due to the low signal strength at energies near the bottom of the gap (few states). As the overall number of states increases with increasing energy, we see an increase in the normalized cross correlation coefficient. The normalized cross correlation then declines and levels out once it has reached the energy where the QPI terminates and where the LDOS kink begins[16], $\Delta_0$. At this point, the QPI stops dispersing and the maximum normalized cross correlation coefficient possible is set by the Gaussian masks ability to reproduce the intensity variations in the data. The lower normalized cross correlation coefficient on the positive side is due to the presence of the pseudogaps $S_{x,y}$ peak[16] which causes the masking to fail. Since this peak has a short wavelength of $\sim 2\ 3/4\ \pi/a_0$ and all the longer wavelength $q$-vectors are gone by this point, the Gaussian mask fails to be able to reproduce the data.

This Gaussian masks failure at high positive energies can be seen clearly in figure 6b, where the Gaussian masks FWHM is plotted as a function of energy. At low energies the FWHM is very high and it rapidly decreases as both positive and negative energies are increased. At high positive energies the FWHM sudden increase is due to the presence of $S_{x,y} \sim 2\ 3/4\ \pi/a_0$. The Gaussian mask also captures the QPI termination phenomena as an increase in the effective length scale. The length scale at the termination energy is approximately equal to that of maps of the local termination energy disorder ($\Delta_0$)[44]. The onset of the local termination energy is shown as a histogram in its values[44] measured for the positive energies of the UD74 K data set in figure 6b. The rapid decrease in the intensity of the short wave length QPI vectors precedes the onset of the local distribution of $\Delta_0$. This could be due to the extended $k$-space scattering sources being influenced by the PAF-zone boundary at much lower energies, which would not be seen in the peak analysis using the octet model. In figure 5, even at very low energies the $k$-space scattering structure curves inward along the PAF-zone boundary. Since the areas in $k$-space that the CMCR maps are extended and not confined to points as in previous analyses, we are able to resolve this effect at low energies for the first time. This shows that the PAF-zone boundary has an influence on the entire energy range of the QPI and may be the cause of the rapid suppression of the QPI peaks intensity.

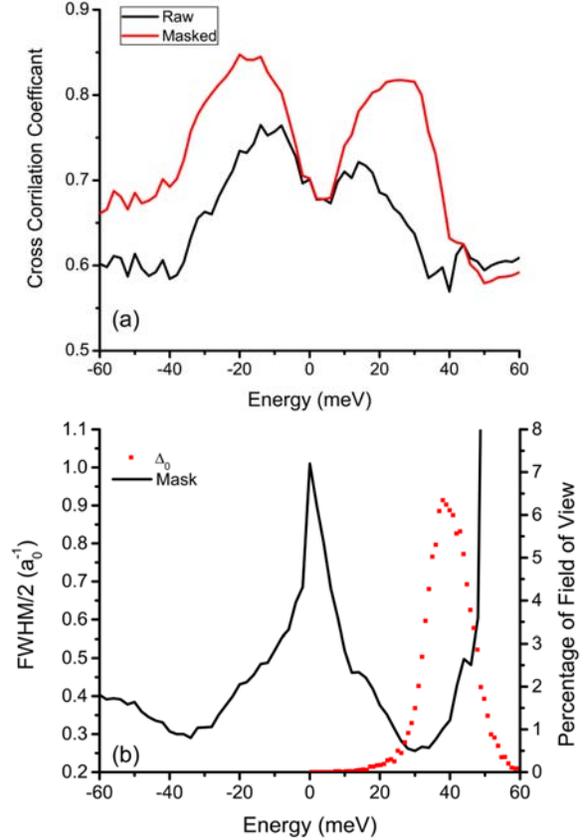

**Figure 6:** a) The normalized cross correlation coefficient between the data and the $q$-space reconstruction for the UD 74 K data. Both the unmasked and the Gaussian masked reconstruction are shown. b) The Gaussian masks FWHM is plotted and the distribution of the real space $\Delta_0$ energy scale is shown as red squares[44]. The FWHM goes to a minimum before the onset of the $\Delta_0$ disorder and reaches a value of $\sim 0.4$ nm which is consistent with the length scale of the real space $\Delta_0$ disorder in this sample (value of $\sim 0.6$ nm).

### $k$-space gap structure:

Figure 7 shows one $k$-space quadrants reconstructed gap structure. The background is the intensity of the reconstruction looking from the anti-node towards the node ($\Gamma$ to M). Overlaid on top of it is the previous octet model $k$-space extracted point's data[18], the corresponding higher harmonic fitted gap and a gap made by fitting the leading edge points to a lowest harmonic d-wave gap. What this figure revels about the previous analysis, is that the octet model extracted points lie inside our extended $k$-space reconstruction, near the center.

At low energies, around zero energy, the $k$-space reconstruction also shows a departure from the d-wave gap. This departure extends to higher energies for positive or empty states verse the negative energies, where the dispersion returns to a d-wave leading edge quicker. This is possibly due

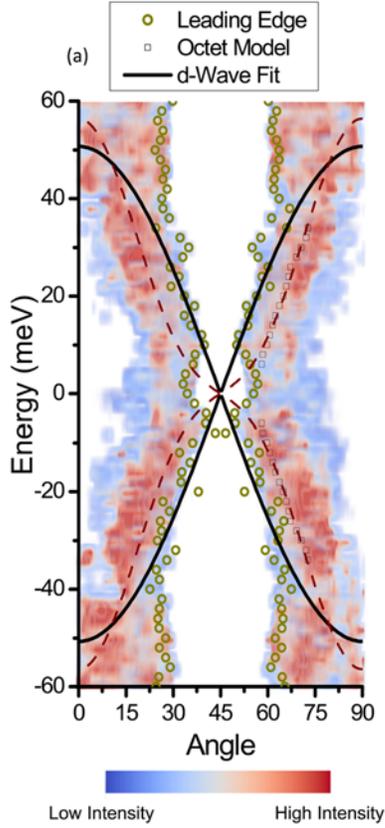

(a)

Legend:
- ○ Leading Edge
- □ Octet Model
- — d-Wave Fit

Low Intensity — High Intensity

**Figure 7:** Looking at the node from the $(\pi,\pi)$ direction the $\mathbf{k}$-space reconstruction is splayed out from -60 meV to 60 meV. The open squares on the right hand side show the $\mathbf{k}$-space positions determined by the previous octet model fitting. This is fit with a dark red line by a d-wave lowest harmonic gap + a higher harmonic $6\theta$ term. The open circles represent the inner edge of the $\mathbf{k}$-space reconstruction; this is fit by a lowest harmonic d-wave gap shown by the solid black line. Near zero energy the $\mathbf{k}$-space reconstruction deviates from this line likely due to unitary scatters.

to the zinc impurities and copper atom vacancies which have resonances at low energies[7] and which in the case of the zinc impurities, have a peak feature on the positive side of the spectrum. It is also possible that the scatters have an s-wave symmetry and therefore do not couple to the d-wave states at low energy[53] causing this departure.

Also of note is that the leading edge of the $\mathbf{k}$-space reconstruction can be fit to a pure d-wave gap up until near the termination energy, where the dispersive structure transitions into a non-dispersive $q_1^*$ structure. This d-wave gap has a lower energy then that seen in SI-STM data, both measured by using the octet model applied to the QPI pattern and as measured from the peak in the LDOS. It is, however, consistent with an ARPES study that shows a full d-wave gap across the entire Fermi surface[17]. This ARPES resolved gap has a value of 42 meV for dopings near UD74 K which matches the value seen here which is 43 meV. What SI-STM is resolving is the effects of the gap structures $\mathbf{k}$-space weight shifting due to the interaction with the PAF-zone boundary and/or an s-wave scattering symmetry matrix term. This observation implies that the wide zero energy arc[18] and the higher harmonic gap seen by SI-STM are due to a PAF-zone boundary or scattering effects which causes the states in $\mathbf{k}$-space that contribute to the QPI pattern fall outside the d-wave gap.

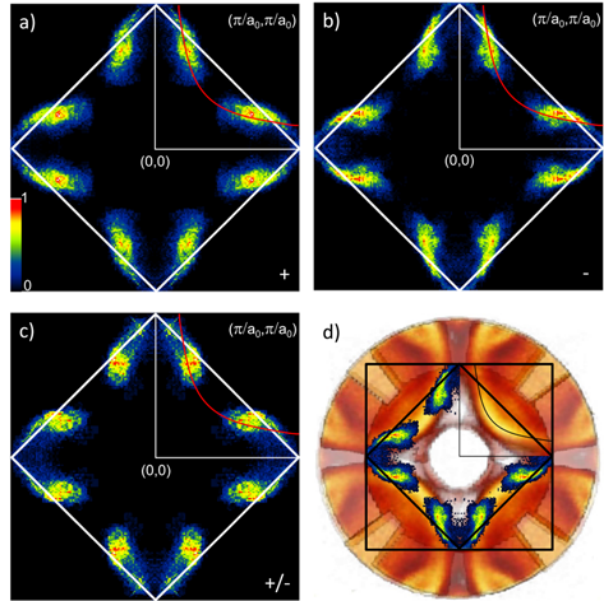

**Figure 8:** The integrated $\mathbf{k}$-space reconstructed density of states from $|0-40$ meV| for the UD74 K sample on the positive side (a), negative side (b) and from the ratio map (c) overlaid with the PAF-zone boundary in white and the fit tight binding Fermi surface in red. The termination at the PAF-zone boundary can be seen as the $\mathbf{k}$-space scattering sites fold over and follow the PAF-zone boundary out towards the anti-node. The lack of observed sites at the node is also evident. In d) the integrated $\mathbf{k}$-space reconstruction is overlaid on top of the integrated and symmetrized ARPES data for a UD76 K lead doped Bi-2212 sample[55]. The first Brillouin zone is shown inside the square, while the PAF-zone boundary is shown imbedded within. Overlaid is our fitted tight binding Fermi surface to the $\mathbf{k}$-space reconstruction, showing excellent agreement with the ARPES data.

### Integrated $\mathbf{k}$-space pattern:

In order to compare the CMCR measured $\mathbf{k}$-space reconstruction to the equivalent ARPES data, and also to display the differences between the positive, negative and the ratio map $\mathbf{k}$-space reconstructions, we show energy integrated $\mathbf{k}$-space reconstructions in figure 8. In the upper right of each data set there is a tight binding fit[54] to the $\mathbf{k}$-space reconstruction. The PAF-zone boundary is plotted as the inscribe square. In 8a and 8b the positive and negative energies are shown respectively, both demonstrating the influence of the PAF-zone boundary. The ratio map, in 8c, shows a similar PAF-zone boundary structure, which means that setup condition artifacts do not cause the inward bending of the $\mathbf{k}$-space recon-struction at the anti-nodes. The **k**-space weight in the ratio map is also shifted from the tight binding fit to the positive and negative energies data. In figure 8d the negative **k**-space reconstruction is overlaid on ARPES data for a lead doped UD76 K sample[55]. The lead removes the supermodulation from the sample and allows a cleaner APRES signal to be measured. The **k**-space reconstruction measured by SI-STM matches that of a similar $T_c$ ARPES sample showing that the two separate measurements of the band structure agree up until the dispersion reaches the PAF-zone boundary.

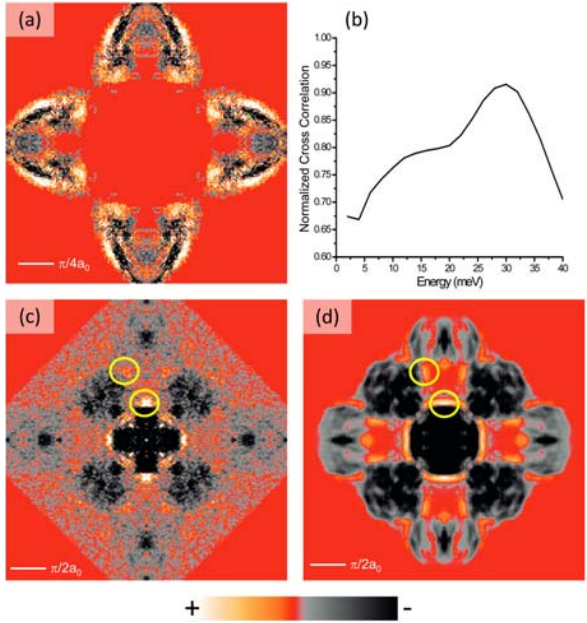

**Figure 9:** a) The difference in integrated **k**-space LDOS intensity between the positive energy side and the negative energy side. Places where the positive side is bigger are represented by the orange to white scale and places where the negative side is bigger are represented by the grey to black scale. b) The normalized cross correlation coefficient between the QPI difference (QPI(+) - QPI(-)) and the reconstructed QPI difference. c) The 20 meV subtracted data layer, displaying prominent QPI peaks that are different between the positive side and negative side. Here in d) we can see these same peaks show up in our reconstructed QPI map. The yellow circles draw attention to the two most prominent differences.

**Positive and negative energy dispersions:**

In order to examine the difference between the positive and negative dispersions, their **k**-space reconstructions can be subtracted and the difference integrated. The positive and negative energy QPI patterns can also be subtracted and the resulting QPI difference maps compared to that predicted by our reconstruction. Figure 9 shows both the positive and negative differences in the **k**-space reconstruction as well as how these differences appear in the **q**-space QPI patterns. Figure 9a is the integrated differences between the positive and negative **k**-space reconstructions. This shows the positive energies disperse towards the anti-node but also disperse out-wards towards $(\pi/a_0, \pi/a_0)$, while the negative energies instead disperse towards the origin. This is expected and can be seen in the theoretical calculated A(**k**,ω) of a d-wave superconductor[32,56]. In figure 9c-d an example of the positive QPI subtracted from the negative QPI at an energy of 20 meV is shown. The biggest differences between the two energies are in the placement of $q_1$, $q_2$, $q_3$ and $q_6$. The shift in these QPI **q**-vectors is highlighted by the yellow circles, both in the data, 9c, and in the **q**-space reconstruction, 9d. Figure 9b shows the normalized cross correlation between the positive - negative QPI data and the positive - negative **q**-space reconstruction. There is excellent agreement across energies and the CMCR reconstruction accurately captures the two different dispersions. This difference cannot be seen in the ratio map and highlights the power of the CMCR method.

**Doping dependence:**

The CMCR **k**-space reconstructions can also resolve doping dependent differences across a wide range of dopings, including at those dopings where the octet model can no longer be applied. At extremely overdoped samples (OD65 K) the QPI pattern no longer has well defined **q**-vectors over a large energy range and it more closely resembles a theoretical A(**k**, ω) since it consists of large continuous **q**-space arcs. This makes it resistant to the traditional octet method of analysis; however the CMCR algorithm can reconstruct this extremely overdoped **k**-space structure. The extremely underdoped samples (UD45 K) have also proven resistant to analysis and the ratio map is needed here in order to remove the large checkerboard signal[18]. However the CMCR algorithm can reconstruct the non-ratio map data at these low dopings as well.

Figure 10a shows three data sets **k**-space reconstructions, OD65 K, UD74 K and UD45 K along with corresponding tight binding fits. The **k**-space structure evolves as a function of doping and at extremely overdoped samples, there is an increase in the weight at the anti-node which overlaps the PAF-zone boundary. For all dopings there is a still a complete lack of signal from the nodal region.

The tight binding fits to the **k**-space reconstructions show an excellent agreement between the Presland phenomenological formula[57] for doping and the Luttinger theorem for a Mott insulators[39] doping (half zone). In figure 10b as the doping is increased and the **k**-space structure shifted towards the origin, there is a convergence in the half-zone and full-zone doping predictions. At these high dopings, the extremely overdoped data set also shows a shift in the length scale as a function of energy (figure 10c). At low energies in the OD65 K data set there is a spike in the length scale followed by a region that has a flat energy dependence. This could be due to a shift in the nature of the scattering at high dopings due to a weakening influence of the PAF-zone boundary. The UD45 K data has a length scale that is similar in energy dependence to the UD74 K data, although it reaches its base value at a lower energy. This more rapid decline is likely due to the increased signal of the checkerboard in comparison to the dispersive QPI in the raw data.

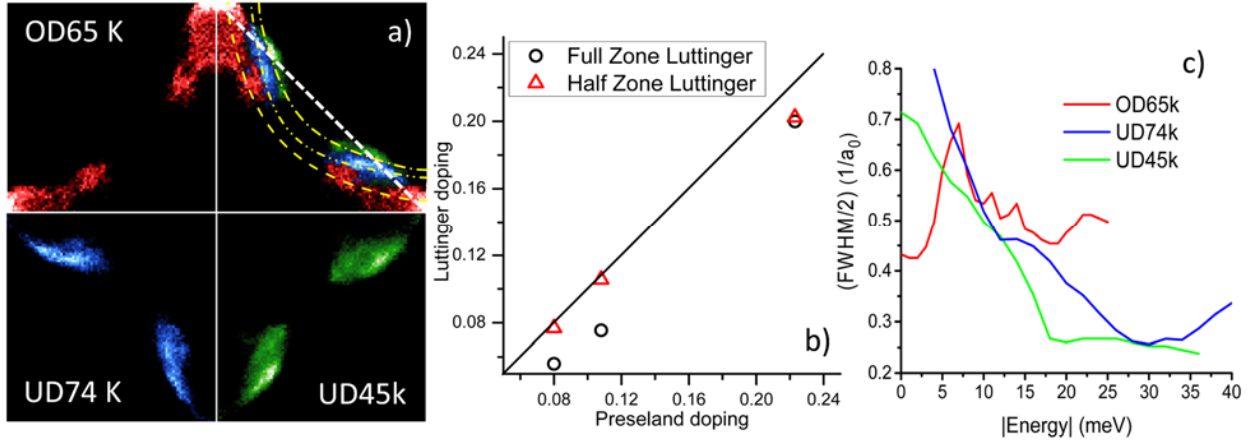

**Figure 10:** The integrated *k*-space reconstructions for three different dopings OD65 K in red, UD74 K in blue and UD45 K in green. In the top right quadrant we have all three overlaid. The progression between dopings driven by chemical potential and band structure changes is evident. Overlaid in yellow are our three tight binding Fermi surface fits to the data. The white dotted line shows the PAF-zone boundary. b) The Presland doping compared with the Luttinger doping for each of the three fitted Fermi surfaces. We see excellent agreement if we use a half zone model. c) The length scale for all three samples, here OD65 K is analyzed up to 25 meV, UD74 K up to 40 meV and UD45 K up to 36 meV. There is a falloff in the FWHM for both the underdoped samples, but the OD65 K sample maintains strong short wavelength QPI out to the LDOS peak energy scale (25 meV).

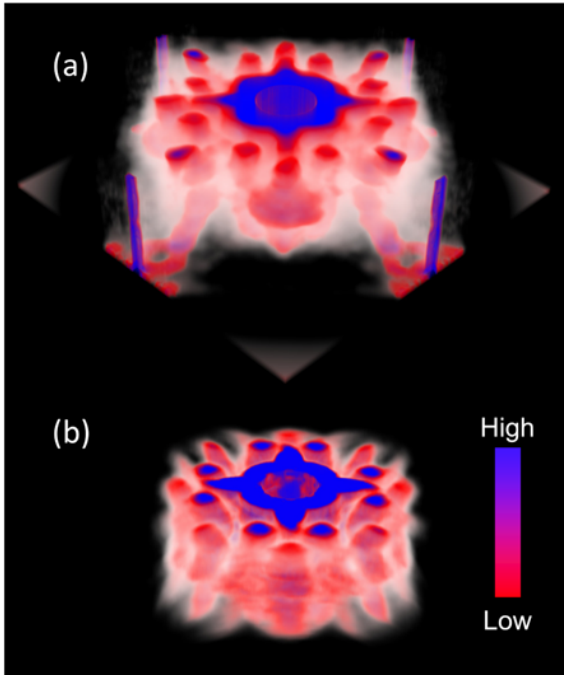

**Figure 11:** QPI (top) and reconstructed QPI (bottom) for UD74 K. In the data the atomic lattice peaks can be seen in the corners as well as the clear difference in the $q_5$ dispersions for positive and negative energies. Positive energies are on the bottom and negative energies on the top.

**Conclusions:**

For the first time the CMCR method allows the detailed *k*-space scattering structure of Bi-2212 to be determined. The resulting reconstruction shows a scattering structure that agrees with ARPES in terms of *k*-space weight up until the PAF-zone boundary. It also agrees with ARPES by showing a pure d-wave energy gap in the underdoped regime that agrees in gap value and form. The CMCR method determines an effective length scale from the wavelength dependent intensity variations of the QPI signal. This length scale decreases monotonically until the QPI terminates. When this happens, the length scale matches the length scale of the real space low energy disorder[44] seen in $\Delta_0$ maps. This $\Delta_0$ energy marks the termination of the V shaped LDOS and the length scale bottoming out in *q*-space occurs at an energy preceding the onset of the real space $\Delta_0$ disorder. The CMCR method also shows the differences between the positive and negative dispersions which is consistent with a JDOS A(*k*,ω) model of a d-wave superconductor. The differences shown between the ratio maps and the positive/negative maps highlight the danger of relying on the ratio map, which conflates the two dispersions and may cause a shift in both the gap structure and the *k*-space points. This is especially problematic if the theory used to interpret the QPI contains these positive and negative differences, such as is the case with a superconducting A(*k*,ω).

The PAF-zone boundaries effects on the energy dispersion of the QPI is mapped in detail for the first time and instead of a simple termination of the dispersive QPI, the PAF-zone boundary is shown to cause the *k*-space scattering origins to curve along it even at low energies. This curvature is resolved using the CMCR method since it measures extended *k*-space regions and it takes into account the shape of the QPI peaks in determining the shape of the *k*-space scattering origins. The large energy range effect of the PAF-zone boundary could also be the cause of the higher harmonic gap structure seen by SI-STM in both the LDOS(E)[16,44] and in the octet models extracted *k*-space gap[4,18]. Figure 11 shows both the reconstructed QPI pattern and the data, highlighting the excellence of our reconstruction, as well as the differences between positive and negative energies in the data. The large bump seen in the middle of the data is the widening of the gap at low energies caused by unitary scatters or scatter symmetry effects (see figure 7).

While the observed dispersive QPI pattern in Bi-2212 has always been associated with Bogoliubov quasiparticles[1,4,18], we should point out that our work here does not assume that the pattern we are inverting comes from Bogoliubov quasiparticles. The CMCR method, due to its generality, will find *k*-space inverse of any *q*-vector (this can also be a drawback). This allows it to track *k*-space scattering sources if they transition from Bogoliubov quasiparticles to pseudogap states[18]. If the pseudogap excitations quasiparticles no longer come from cooper pairs, then this transition could be the source of the change in our observed length scale. This transition is also consistent with past claims about the nature of the pseudogap vs. the dispersive QPI[18].

The CMCR method provides the unique ability to determine the *k*-space scattering structure from complex QPI patterns. This allows unknown band structures to be measured and detailed information about how the scattering is affected by other forms of order in the system to be resolved. The ability to tune the model and in the future to add matrix elements, multiple bands and other elements to it, makes it an excellent tool for understanding the scattering and band structure of a crystal from SI-STM data. The methods automation removes much of the guess work that was formerly involved and it will enable the inversion of complex QPI patterns that are currently undergoing investigation. As shown here, applying it to well-known and studied QPI patterns, can be quite illuminating and allows their scattering to be studied in far greater detail than was previously capable. The CMCR method represents a powerful new technique that will greatly expand the capabilities of SI-STM by allowing the determination of *k*-space scattering structures, the determination of hidden scattering information and in the future it will enable the measurement of unknown band structures from high quality SI-STM data.

**Acknowledgements:**

We would like to acknowledge the support of J.C. Davis and John Moreland, without which this paper would not have been possible. Also thanks to Jinho Lee for his role in inspiring the project. Support for this work was provided by the Sloan Foundation. We would also like to thank Nvidia for their hardware grant program, which allowed the acceleration of the code and its completion in a reasonable time frame.